\documentclass{article}

\PassOptionsToPackage{numbers, compress}{natbib}

\usepackage[preprint]{neurips_2021}

\usepackage[utf8]{inputenc} %
\usepackage[T1]{fontenc}    %
\usepackage{hyperref}       %
\usepackage{url}            %
\usepackage{booktabs}       %
\usepackage{amsfonts}       %
\usepackage{nicefrac}       %
\usepackage{microtype}      %
\usepackage{xcolor}         %
\usepackage{amsmath}        %
\usepackage{bm}
\usepackage{multirow}
\usepackage{extarrows}%
\usepackage{subcaption}
\usepackage{multicol}
\usepackage{bbding}

\usepackage{graphicx}       %
\def\R{\mathbb{R}}
\DeclareMathSymbol{:}{\mathord}{operators}{"3A}

\title{Improving On-Screen Sound Separation for Open-Domain Videos with Audio-Visual Self-Attention}
\author{%
  Efthymios Tzinis\thanks{Work done during an internship at Google.} \\
  UIUC\\
  \texttt{etzinis2@illinois.edu} \\
   \And
  Scott Wisdom \\
  Google Research\\
  \texttt{scottwisdom@google.com} \\
  \AND
  Tal Remez \\
  Google Research\\
  \texttt{talremez@google.com} \\
  \And
  John R. Hershey \\
  Google Research\\
  \texttt{johnhershey@google.com} \\
}
\date{Last compiled \today}

\begin{document}

\maketitle

\begin{abstract}
We introduce a state-of-the-art audio-visual on-screen sound separation system which is capable of learning to separate sounds and associate them with on-screen objects by looking at in-the-wild videos. We identify limitations of previous work on audio-visual on-screen sound separation, including the simplicity and coarse resolution of spatio-temporal attention, and poor convergence of the audio separation model. Our proposed model addresses these issues using cross-modal and self-attention modules that capture audio-visual dependencies at a finer resolution over time, and by unsupervised pre-training of audio separation model. These improvements allow the model to generalize to a much wider set of unseen videos. We also propose a robust way to further improve the generalization capability of our models by calibrating the probabilities of our audio-visual on-screen classifier, using only a small amount of in-domain videos labeled for their on-screen presence. For evaluation and semi-supervised training, we collected human annotations of on-screen audio from a large database of in-the-wild videos (YFCC100m). Our results show marked improvements in on-screen separation performance, in more general conditions than previous methods.
\end{abstract}

\section{Introduction}
Humans are able to effortlessly perceive sounds in a noisy scene, and associate them with any corresponding visible objects.  In audio processing, a corresponding challenge is to isolate sound sources from a mixture waveform and identify the associated visual appearance of each sound source. Sound sources in audio are analogous to objects in computer vision.  An advantage in vision is that objects tend to occupy distinct regions of pixels, so that features from different regions of the image correspond to different objects. In contrast, sound sources are superimposed in the audio, making it even more difficult to disentangle the features of different sounds.  

This creates a challenge for unsupervised learning of audio-visual separation, because unlike their visual counterparts, the audio sources in a scene cannot be easily selected for alignment with the video objects.  The audio needs to be separated into sources at some point in the process, either by conditioning separation on selected video objects, or by separating the audio \emph{a priori}, before associating the sounds with video objects, or both.   The \emph{a priori} separation of the sounds, which we pursue here, has a few advantages: thanks to recent work, it can be learned in an unsupervised way, and it can handle an unknown number of sounds, including those that do not appear on-screen.   

Despite the recent progress in audio-visual machine perception using neural networks, trained models still suffer from a variety of problems.  Many of the previous works rely heavily on labeled training data, or data with special properties such as having one high quality sound that is on screen in each example. 
They may also rely on pre-trained supervised object detectors that require labeled data.  This can work well in a restricted domain, where such labeled data are available; however in an open-domain setting the reliance on human labels effectively precludes scaling to large open-domain data.  Thus a typical problem is difficulty in generalizing from the domain of labeled training data to the domain of realistic data that would be seen in a practical application.

Although the co-perception and alignment of audio and visual modalities is not trivial, a variety of works have shown promising results by proposing neural network architectures apt for the task \cite{cheng2020lookListenAttendSSLAVRepresentationLearning, wu2019dualAttentionForAVEventLocalization, bertasius2021divisibleAttention, afouras2020selfsupAVObjectsOpticalFlow, gao2019coSeparation, tzinis2021into, chatterjee2021visual}.  Audio-visual sound separation \cite{hershey2002audio, ephrat2018looking, afouras2018conversationDeepAVSpeechEnhancement} and more specifically separation of on-screen versus off-screen sounds \cite{owens2018audioFirstOnScreen} have also seen remarkable performance improvements since the initial works.  Important innovations have included using localization of objects \cite{wu2019dualAttentionForAVEventLocalization, zhu2021visually_guided_SS_localization,  lin2021unsupervisedSoundLocalizationIterativeContrastive}, forcing consistency between audio and visual modalities representation \cite{lee2021CrossModalAffinityAudVisSpeechSeparation, gao2021visualvoiceSpeechSeparationCrossModalConsistency, arandjelovic2018objectsThatSound, gao2019coSeparation}, weakly \cite{rahman2021weaklySupervisedAVSoundSourceSeparation} and self-supervised approaches \cite{rouditchenko2019selfAVCo_Segmentation, korbar2018cooperativeSelfSupervisedSynch, afouras2020selfsupAVObjectsOpticalFlow, tzinis2021into}. Despite the remarkable progress in the field of on-screen sound source separation, most of these works are constrained to isolating a specific set of sounds which can appear on-screen such as speech \cite{ephrat2018looking, gao2021visualvoiceSpeechSeparationCrossModalConsistency, afouras2018conversationDeepAVSpeechEnhancement} or music \cite{gan2020musicGestureforAVSeparation, gao2018learningToSeparateObjectFromUnlabeledVideo}. Recent work has started to expand beyond music and speech to a wider variety of classes, using visual scene graphs to model relationships between detected objects and use this information to condition audio separation \cite{chatterjee2021visual}, but this approach is still inherently limited by requiring labeled data to train supervised object detection. Although the seminal works in separating on-screen sounds proposed models which were somewhat invariant to the types of sources \cite{owens2018audioFirstOnScreen, gao2018learningToSeparateObjectFromUnlabeledVideo}, those systems were unable to be trained with real world videos mainly because they needed labeled videos assuming that sound sources always appeared on-screen during training.  A recently proposed model, known as AudioScope \cite{tzinis2021into}, addressed this problem by enabling unsupervised training of audio-visual sound separation models from in-the-wild videos, without requiring object detection modules nor assuming that all sources have to be on-screen.  The model is trained to separate individual sound sources from a mixture of two videos using mixture invariant training (MixIT) \cite{MixITNeurIPS}, which works by assigning estimated sources to the best matching video mixture. The AudioScope model then uses these source assignments as pseudo-labels to help train the audio-visual correspondence classifier.  

However, the problem of generalizing to a wide variety of videos still remains largely unsolved. AudioScope \cite{tzinis2021into} was able to obtain adequate performance in terms of on-screen sound detection and reconstruction when training was injected with a few thousands labeled examples, but performance suffered in the purely unsupervised setting. In this work, we try to identify some of the possible limitations of this approach and propose ways to overcome the associated challenges. 
We hypothesize that AudioScope's performance is limited by the simplicity of its visually guided spatio-temporal attention layer \cite{BahdanauCB14}, and the low temporal resolution (one frame per second) of its visual model. These factors may prevent AudioScope from capturing synchronization features which might be crucial for detecting the inter-play between audio and video \cite{hershey2000audio, korbar2018cooperativeSelfSupervisedSynch, afouras2020selfsupAVObjectsOpticalFlow}. Another limitation of the AudioScope architecture stems from the outputs of the audio-visual classifier which are produced independent of the possible domain mismatch between the train/test distributions. Consequently, the probabilities which are later used as soft weights in front of the separated sources might be heavily biased because of slight variations in the recording setup (e.g. audio-energy levels, camera position, etc.), thus, leading to poor generalization to previously unseen audio and video classes.  

We propose to address these problems by introducing richer transformer-style audio-visual attention models, operating at higher frame rates, in order to better detect audio-visual synchrony.  Furthermore, we propose to refine the estimates of the on-screen audio-visual classifier per source by utilizing the powerful representations obtained from unsupervised pre-training of the audio source separation module with MixIT. We also propose a new way to perform weakly-supervised domain adaptation by calibrating the output probabilities produced by the classifier using a very small amount of videos which were annotated for their on-screen presence. Finally, we also propose even more challenging evaluation datasets containing unfiltered videos from the YFCC100m \cite{thomee2016yfcc100m} data collection, and show that our new proposed models both generalize and perform better on these evaluations.

\section{Relation to Prior Work}
Audio-visual separation of in-the-wild on-screen sounds relies on the capability of separating the individual sources that are superimposed in an input audio recording. Recent work has shown that it is possible to train an open-domain model to isolate individual sound sources using a mask-based convolutional architecture regardless of the category of sound \cite{kavalerov2019universal, tzinis2020two}. 
A related promising direction is to extract sources of interest by conditioning the separation networks using identity cues.  This has yielded performance improvements for speech \cite{wang2019voicefilter} as well as universal source separation \cite{tzinis2020improving, ochiai2020listenToWhatYouWant}. 
However, these experiments relied on having sufficient supervised training data and were evaluated only on test sets with similar environmental conditions and sound distributions.  In order to extend the reach of this approach, methods have been proposed to train separation models with no access to ground truth clean sources by utilizing weak class labels \cite{pishdadian2019finding}, the spatial separability of the sources \cite{tzinis2019unsupervised,seetharaman2019bootstrapping,drude2019unsupervised} and self-supervision in the form of MixIT \cite{MixITNeurIPS}. This makes it possible to learn separation of signals well outside the domains for which isolated source databases exist.  

AudioScope \cite{tzinis2021into} used MixIT in order to avoid the dependency on clean in-domain training data, which may be unavailable for many types of sounds. However, it also relied on filtering of the dataset in order to increase the proportion of videos containing on-screen sounds.  This was done using scores produced by an audio-visual coincidence model \cite{jansen2020coincidence} pre-trained  on AudioSet \cite{AudioSet}.  A disadvantage of this approach is that the coincidence model may be susceptible to a looser semantic association, where sounds are thematically related to a visual scene, as opposed to true audio-visual correspondence, where the sounds actually appear onscreen.  This, and any other biases in that model, may have limited the ability of the resulting AudioScope model to generalize to the distribution of sounds in unfiltered data.  Our approach extends the utilization of MixIT by pre-training on audio from a broader in-the-wild video data collection.  This may enable less reliance on pre-filtering, perhaps by producing more robust pseudo-labels for training the audio-visual classifier. Building upon that, we propose a novel way to refine the on-screen probability estimates and perform domain adaptation on an on-screen separation model using isotonic regression \cite{zadrozny2002transforming} with a small subset of labeled videos for their on-screen presence. 

An open question in audio-visual correspondence models concerns the level of processing at which audio and video objects can be aligned.  Typically audio-visual models have used high-level features at the output of deep neural networks to estimate correspondence between audio and video signals \cite{lee2021CrossModalAffinityAudVisSpeechSeparation,tzinis2021into,jansen2020coincidence,korbar2018cooperativeSelfSupervisedSynch,chatterjee2021visual}. Such high-level representations may tend to focus on semantic information about the class of objects and sounds, especially when the features are computed from single frames at a low frame rate.  Such methods may work well for single instances of a class of object or sound, but may struggle with identification for multiple instances of a class, or for classes not seen during training.   
In contrast, there may be significant information in the correspondence between lower-level features. Mutual information between low-level features was used for audio-visual localization \cite{hershey2000audio}, and several more recent works have shown promising results for self-supervised audio-visual learning using low-level, motion \cite{zhu2021visually_guided_SS_localization} and optical flow \cite{afouras2020selfsupAVObjectsOpticalFlow} features. Such features may help with generalization and instance-level correspondence by detecting synchronous dynamics of the audio and video, regardless of their semantic class.

Attention is a framework that may be useful at multiple levels of processing. Attention-based models have been extensively utilized across audio-visual learning tasks. An attention-based framework was recently used to modulate audio representations using motion-based visual features \cite{zhu2021visually_guided_SS_localization} for separation and localization. Conversely, modulating video features based on audio embeddings has also been used for speech separation \cite{lee2021CrossModalAffinityAudVisSpeechSeparation} as well as in the spatio-temporal attention module of the initial \textit{AudioScope} \cite{tzinis2021into}. Other works combined self-attention layers \cite{vaswani2017attention} for modeling inter-modality temporal patterns as well as cross-modal attention modules for intra-modality associations \cite{yu2021mpnSA_CMA_EventLocalization, cheng2020lookListenAttendSSLAVRepresentationLearning, wu2019dualAttentionForAVEventLocalization} for sound source localization.  Our proposed model in this paper applies attention at the level of audio sources, which is similar to the operation of \emph{slot attention} models which apply attention across video objects \cite{locatello2020object}.  

Inspired by the flexibility of self-attention models to represent spatio-temporal associations, we propose to extend the applicability of self-attention and cross-modal attention modules to unsupervised audio-visual on-screen sound separation tasks. However, the applicability of these attention layers is not trivial when we also want to scale to higher resolution audio-visual representations in order to capture synchrony. The attention tensor grows quadratically in  complexity with the dimensionality of the space. For this reason, we propose a variety of self-attention architectures that factorize the attention operation across different dimensions and modalities. The separable attention layers allow us to achieve similar performance to full self-attention with a much lower computational footprint. Our proposed layers differ from recently proposed separable attention layers \cite{bertasius2021divisibleAttention, li2021vidtrSeparableAttentionVideo} in the sense that we also accommodate modeling intra-modality patterns from both audio and video features.

\section{Model architecture}
\label{model}
\begin{figure}[ht]
  \centering
  \includegraphics[width=1.0\linewidth]{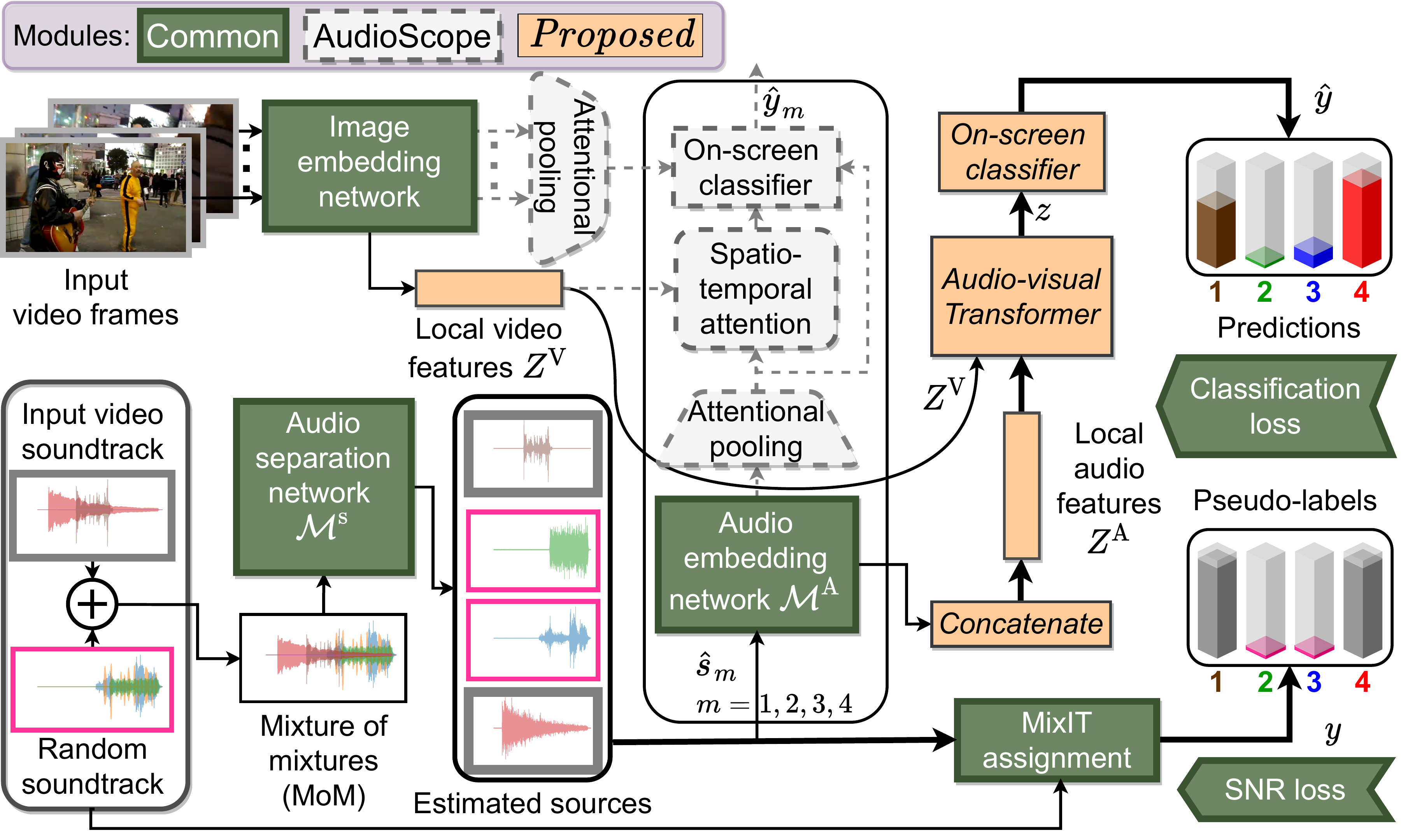}
\caption{System diagram comparing our proposed approach with AudioScope \cite{tzinis2021into}, with two input mixtures and four output sources. The modules common between AudioScope and our approach are represented with green dark background, AudioScope modules that are replaced use lighter font and dashed border, and new proposed modules have orange background with solid border.}
  \label{fig:schematic}
\end{figure}

In Figure \ref{fig:schematic}, the overall architecture of the proposed model 
is displayed alongside the replaced modules from the previous version. The most noticeable changes are with respect to the way we align the audio-visual features extracted from the embedding networks. In the proposed version, we introduce a family of transformer-based architectures (explained in Section \ref{model:attention}) which are able to obtain semantically rich representations between the audio and the video modalities. In addition, we omit the temporal pooling operation for the inputs of the proposed attention blocks since we want to let the model build implicit latent representations capable of capturing low-level intra-modal synchrony-based dependencies.    

\subsection{Separation Module}
\label{model:separation}
The separation module $\mathcal{M}_s$ has the same dilated convolutional architecture as the one used in AudioScope \cite{tzinis2021into} with learnable encoder and decoder. This module takes as input a mixture waveform $x \in \R^{T'}$, estimates $M$ masks in the encoded latent space and as a result outputs $M$ estimated sources $\hat{s} \in \mathbb{R}^{M \times T'}$ which are forced to add up to the input mixture through a consistency layer \cite{wisdom2018consistency}. Based on the ablation studies of \cite{tzinis2021into} and our experiments, we completely remove the conditional separation \cite{tzinis2020improving}, as it did not provide any significant gains over not using the visual representations as inputs to the separation framework. More importantly, by removing the dependence on the visual conditional embeddings, we are able to pre-train the separation module on all YFCC100m \cite{thomee2016yfcc100m} audio tracks in order to provide a better initialization for training the audio-visual model.

\subsection{Audio and video embedding networks}
\label{model:embedding_networks}
We extract the audio features for the $M$ estimated sources $\hat{s}$ and the corresponding $T$ input video frames, with $128 \times 128$ spatial resolution each, using a MobileNetV1 architecture~\cite{howard2017mobilenets} following a similar setup used for AudioScope \cite{tzinis2021into}. Specifically, the audio embedding network takes as input log mel-scale spectrogram representations of the time-domain separated sources $\hat{s}$ while the visual embedding network is applied to each of the $T$ input video frames independently. We use local features extracted from intermediate levels of the embedding networks as audio and video features. For the audio part, we use the output of the intermediate activation at the $23$rd layer, while for the video part, we use the one that has $8 \times 8$ spatial locations. These activations provide audio-visual representations which enjoy both local and global feature descriptors. The final audio and video feature tensors are also fed through dense layers in order to force them to have the same depth $D$.

\begin{figure}[!htb]
    \centering
    \begin{subfigure}[h]{0.3272\linewidth}
      \includegraphics[width=\linewidth]{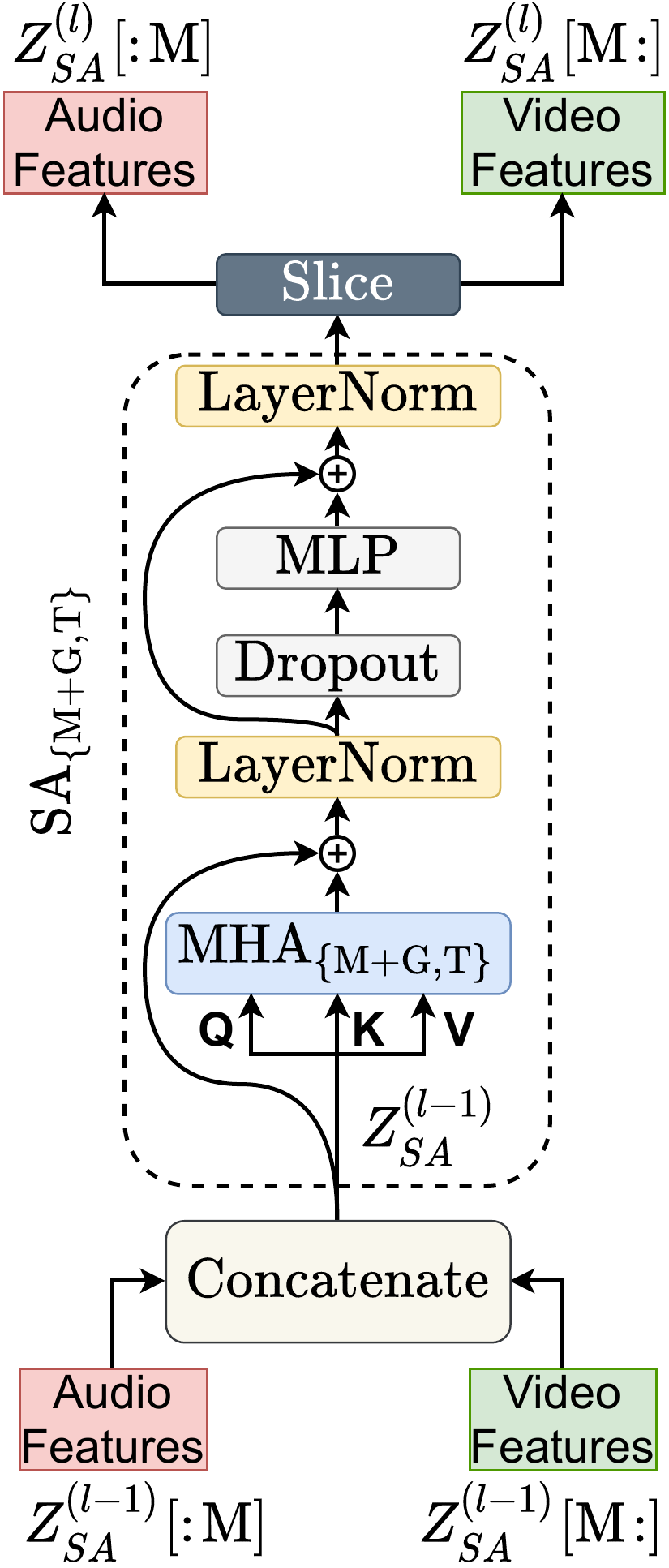}
      \caption{Self-Attention (SA).}
      \label{fig:sa_block} 
     \end{subfigure}
     \begin{subfigure}[h]{0.437\linewidth}
      \includegraphics[width=\linewidth]{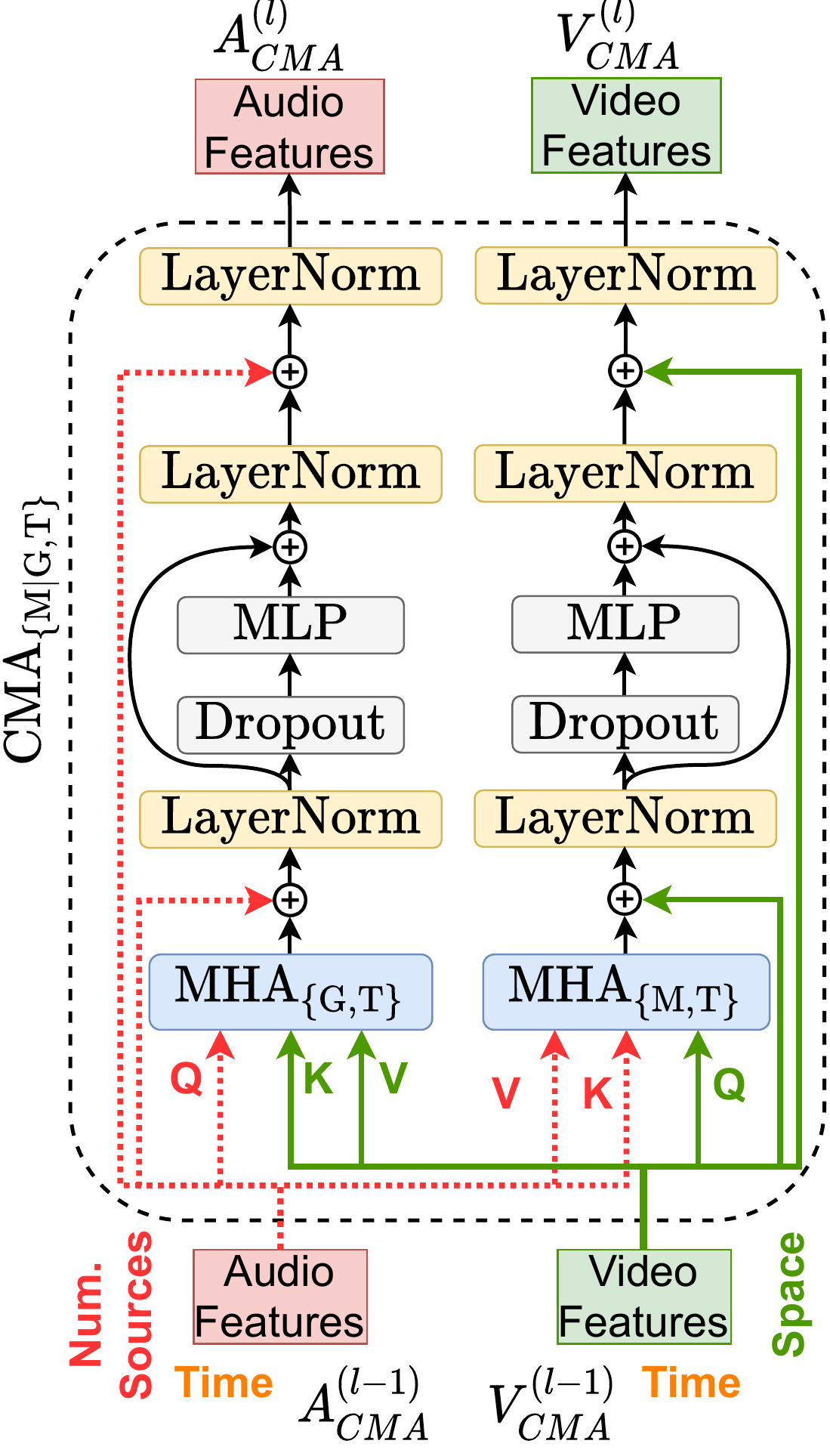}
      \caption{Cross-modal attention (CMA).}
      \label{fig:cma_block} 
     \end{subfigure}
  \begin{subfigure}[h]{0.21725\linewidth}
      \includegraphics[width=\linewidth]{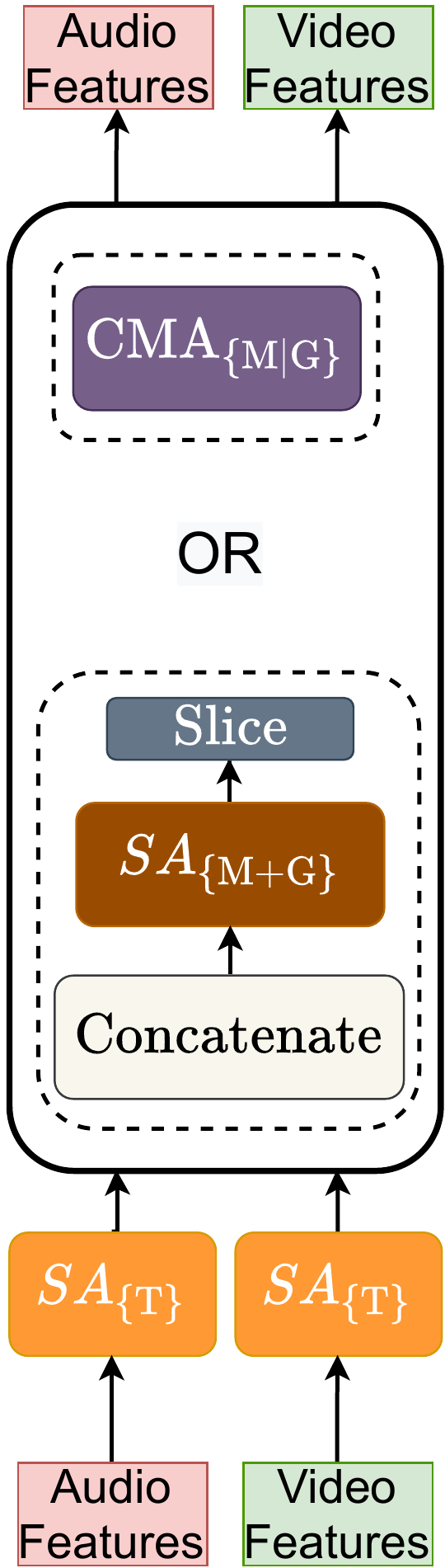}  
      \caption{Separable attention.}
      \label{fig:separable_cma_block} 
     \end{subfigure}\\
    \caption{Attention architectures for audio-visual alignment and feature extraction. All features are depicted without including the batch and depth dimensions for simplicity.}
    \label{fig:attention_architectures}
\end{figure} 

\subsection{Audio-visual spatio-temporal attention}
\label{model:attention}
We seek to propose mechanisms which are able to effectively capture correlation across source, space, and time for the audio features $Z_\mathrm{A} \in \R^{M \times T \times D}$ and the corresponding video features $Z_\mathrm{V} \in \R^{G \times T \times D}$, where $G=8\cdot8$ is the total number of spatial locations, and the time dimension $T$ is shared across both tensors. We provide a slightly more general version of an attention layer \cite{BahdanauCB14} which computes similarities between a packed tensor of queries $Q \in \R^{X_Q \times T_Q \times D}$ with respect to some packed keys $K \in \R^{X_K \times T_K \times D}$, where $D$ denotes the depth dimensionality of the tensors. The similarities are computed using a generalized version of the typical inner product for tensors $\langle Z_1, Z_2 \rangle_{\mathcal{A}}$, which reduces across the specified dimensions $\mathcal{A}$ of the second tensor $Z_2$. Note that we assume that the dimensions of $Z_2$ are a subset of the dimensions of $Z_1$. 
By using a scaled tensor inner-product \cite{vaswani2017attention} and a $\operatorname{softmax}_\mathcal{A}$ activation that averages over the dimensions specified by $\mathcal{A}$ at the output of the tensor product $\langle K, Q\rangle_{\{\mathrm{D}\}}$ we produce the resulting similarity tensor which modulates the values $V \in \R^{X_V \times T_V \times D}$:
\begin{equation}
    \begin{aligned}
        \operatorname{attention}_{\mathcal{A}}(Q, K, V) = 
        \langle \alpha, f_{\mathrm{V}}(V)\rangle_{\mathcal{A}}, 
        \enskip 
        \alpha = \operatorname{softmax}_\mathcal{A}\left(\frac{1}{\sqrt{D}} 
        \langle f_{\mathrm{K}}(K), f_{\mathrm{Q}}(Q) \rangle_{\{\mathrm{D}\}}
        \right),
    \end{aligned}
    \label{eq:general_attention}
\end{equation}
where $Q$, $K$, $V$, and $\alpha$ are the query tensor, the key tensor, the value tensor, and the attention weight distribution tensor across the set of specified axes $\mathcal{A}$ of the value/key tensors over which attention is applied. For example, for input query $Q$ of shape $X_Q\times T_Q \times D$ and value $V$ of shape $X_V\times T_V \times D$, $\operatorname{attention}_{\{\mathrm{X_V}, \mathrm{T_V}\}}(Q, V, V)$ performs attention over the first and second axes of $V$, yielding an output tensor of shape $X_Q\times T_Q \times D$. %
The dense layers $f_{\mathrm{Q}}$,  $f_\mathrm{V}$, $f_{\mathrm{K}}$ are all trainable and applied to the depth dimension $D$.

We utilize as a main block of our network the multi-head attention (MHA) layer proposed in \cite{vaswani2017attention}. Each one of the $H$ heads performs attention to some low-dimensional embeddings derived from the tensors $Q$ and $V$, with the output embedding depth reduced to
$D / H$. These independent attention heads have the capability to focus on different semantics of the input tensors. The final output, after performing attention across the specified axes $\mathcal{A}$, is given by aggregating all the intermediate heads' outputs as shown next: 
\begin{equation}
    \begin{aligned}
        o^{(h)} = \operatorname{attention}_{\mathcal{A}}(f_Q^{(h)}(Q), f_V^{(h)}(V), f_V^{(h)}(V)), \enskip h \in {1, \dots, H} \\
        \operatorname{MHA}_{\mathcal{A}}(Q, V) = f(\operatorname{Concat}([o_1, \dots, o_H])) \in \R^{X_Q \times T_Q \times D}
    \end{aligned}
    \label{eq:mha}
\end{equation}
where 
$f$ 
denotes a dense layer 
$\R^{X_Q \times T_Q \times D}
\rightarrow 
\R^{X_Q \times T_Q \times D}$ 
and the dense layers 
$f_Q^{(h)}$
and 
$f_V^{(h)}$
are defined as linear maps: 
$\R^{X_Q \times T_Q \times D} 
\rightarrow  
\R^{X_Q \times T_Q \times D / H}$
and 
$\R^{X_V \times T_V \times D} 
\rightarrow  
\R^{X_V \times T_V \times D / H}$, respectively. 
In this version we have simplified the usage of the attention layer in that we always assume that the keys and values tensors are the same. Now, we describe in detail the proposed versions of the transformer-based \cite{vaswani2017attention} audio-visual attention architectures.

\subsubsection{Self-attention (SA)}
\label{model:attention:sa}
First, we concatenate the audio and the video tensors across the sources and spatial dimensions:
\begin{equation}
    \begin{aligned}
        Z_\mathrm{AV}^{(0)} = \operatorname{Concat}([Z_\mathrm{A}, Z_\mathrm{V}]) \in \R^{(M+G) \times T \times D}
    \end{aligned}
    \label{eq:sa_concat}
\end{equation}
which is used as the input to the first self-attention layer. 

\textbf{Joint}: In the joint version of the self-attention module we seek to model correlations across space, time, and sources by attending across all those dimensions. Formally, we express the operation performed at the $l$-th layer of a joint self-attention module as follows, also illustrated in Figure \ref{fig:attention_architectures}(a):
\begin{equation}
    \begin{aligned}
        r^{(l)} =  & \operatorname{LayerNorm}\left( \operatorname{MHA}_{\{\mathrm{M+G},\mathrm{T}\}}\left(Z_\mathrm{AV}^{(l-1)}, Z_\mathrm{AV}^{(l-1)}\right) + Z_\mathrm{AV}^{(l-1)} \right) \\ 
        Z_\mathrm{AV}^{(l)} = & \operatorname{LayerNorm}\left( f^{(l)} \left( \operatorname{Dropout}(r^{(l)}) \right) \right) + r^{(l)} 
        \end{aligned}
    \label{eq:joint_sa}
\end{equation}
where $f^{(l)}$ is an output dense layer, $\operatorname{LayerNorm}$ is layer normalization \cite{ba2016layernorm} and $\operatorname{Dropout}$ denotes a dropout layer \cite{srivastava2014dropout}. We define the sequence of operations in Equation (\ref{eq:joint_sa}) as $Z_\mathrm{AV}^{(l)} = \operatorname{SA}_{\{\mathrm{M+G},\mathrm{T}\}}(Z_\mathrm{AV}^{(l-1)})$ where the self-attention is performed across the joint sources-and-spatial dimension $\mathrm{M+G}$ and the time axis $\mathrm{T}$. The final representation $z$, after the repetition of $L$ self-attention blocks, for all $M$ sources, is obtained through the appropriate slicing and by performing attentional pooling \cite{tzinis2021into} across the time axis.
\begin{equation}
    \begin{aligned}
    z = & \operatorname{MHA}_{\{\mathrm{T}\}}\left(\sum_t^{T}\widehat{z}_t, \widehat{z}\right) \in \R^{M \times D}, \enskip \widehat{z} = & Z_\mathrm{AV}^{(L)}[:M] \in \R^{M \times T \times D}.
    \end{aligned}
    \label{eq:final_jointselfattetnion}
\end{equation}

\textbf{Separable}: An issue with the joint attention defined above is that as the number of audio sources $M$, spatial locations $G$, and/or time resolution $T$ of the input tensor $Z_\mathrm{AV}^{(0)}$ is increased, the computational and space complexity of the conventional multi-head attention layer grows quadratically. For this reason, we propose a separable version of attention: first attend across time only, $\{\mathrm{T}\}$, and then subsequently attend across the sources and spatial axis, $\{\mathrm{M+G}\}$, similar to a previous work \cite{bertasius2021divisibleAttention}. In this way, we are able to separate the computation and not construct the full attention tensor with space complexity $\mathcal{O}(T^2 \cdot (M+G)^2 \cdot H)$, where $H$ is the number of attention heads. Formally, we write the following sequential operations for the $l$-th layer of the separable self-attention block using the self-attention module defined in Equation (\ref{eq:joint_sa}):
\begin{equation}
    \begin{aligned}
        a^{(l)} = & \operatorname{SA}_{\{\mathrm{T}\}}(Z_\mathrm{AV}^{(l-1)}[1:M]) \\
        v^{(l)} = & \operatorname{SA}_{\{\mathrm{T}\}}(Z_\mathrm{AV}^{(l-1)}[M:M+G]) \\
        Z_\mathrm{AV}^{(l)} = & \operatorname{Concat}(
            [a^{(l)}, v^{(l)}]
        ) \\
        Z_\mathrm{AV}^{(l)} = & \operatorname{SA}_{\{\mathrm{M+G}\}}\left(  Z_\mathrm{AV}^{(l)},  Z_\mathrm{AV}^{(l)} \right).
    \end{aligned}
    \label{eq:separable_sa}
\end{equation}
The final audio-visual representation is obtained through attentional pooling and slicing as before. This is illustrated in Figure \ref{fig:attention_architectures}(c).

\subsection{Cross-modal attention (CMA)}
\label{model:attention:cma}
In this variation of the audio-visual attention module we keep the audio and the video modality tensors separate, and we perform queries from one modality to another. Formally the input to the stacked CMA blocks are always a pair of an audio feature tensor $A^{(0)}_\mathrm{CMA} \in \R^{M \times T \times D}$ and a video feature tensor $V^{(0)}_\mathrm{CMA} \in \R^{G \times T \times D}$.

\textbf{Joint}: We perform a directional attention from the audio (video) modality tensor to the video (audio) tensor, attending across both sources and time (space and time) axes. Formally, at the $l$-th layer we have the following sequence of operations, also illustrated in Figure \ref{fig:attention_architectures}(b):
\begin{equation}
    \begin{aligned}
        a_1^{(l)} = & \operatorname{MHA}_{\{\mathrm{G}, \mathrm{T}\}}\left(A_\mathrm{CMA}^{(l-1)}, V_\mathrm{CMA}^{(l-1)}\right) &
        v_1^{(l)} = & \operatorname{MHA}_{\{\mathrm{M}, \mathrm{T}\}}\left(V_\mathrm{CMA}^{(l-1)}, A_\mathrm{CMA}^{(l-1)}\right) \\
        a_2^{(l)} = & \operatorname{LayerNorm}(a_1^{(l)} + A_\mathrm{CMA}^{(l-1)}) &
        v_2^{(l)} = & \operatorname{LayerNorm} \left(v_1^{(l)} + V_\mathrm{CMA}^{(l-1)} \right) \\
        a_3^{(l)} = & f \left( \operatorname{Dropout}(a_2^{(l)}) \right) & v_3^{(l)} = & f \left( \operatorname{Dropout}(v_2^{(l)}) \right) \\
        A_\mathrm{CMA}^{(l)} = & \operatorname{LayerNorm} \left(a_3^{(l)} + A_\mathrm{CMA}^{(l-1)} \right) & V_\mathrm{CMA}^{(l)} = & \operatorname{LayerNorm} \left(v_3^{(l)} + V_\mathrm{CMA}^{(l-1)} \right). \\
    \end{aligned}
    \label{eq:joint_cma}
\end{equation}
The left side of the set of the above equations describes video-guided attention where we use the video features $v^{(l)}$ to modulate the audio features $a^{(l)}$, and vice versa on the right side. We define the sequence of operations in Equation (\ref{eq:joint_cma}) as $A_\mathrm{CMA}^{(l)}, V_\mathrm{CMA}^{(l)}
= \operatorname{CMA}_{\{\mathrm{M|G}, \mathrm{T}\}}
(A_\mathrm{CMA}^{(l-1)}, V_\mathrm{CMA}^{(l-1)})$ where each cross-modal attention is performed across the dimension of audio sources $M$ or spatial locations $G$ (denoted in our notation as "$\mathrm{M|G}$") and the time axis. The output audio-visual embedding $z$ contains information for all $M$ sources and is obtained after the repetition of $L$ cross-modal attention blocks by performing attentional pooling across the time axis on the output audio tensor, as follows:
\begin{equation}
    \begin{aligned}
        z = & \operatorname{MHA}_{\{\mathrm{T}\}}\left(\sum_t^{T}\widehat{z}_t, \widehat{z}\right) \in \R^{M \times D}, \enskip \widehat{z} = & A_\mathrm{CMA}^{(l)} \in \R^{M \times T \times D}.
    \end{aligned}
    \label{eq:final_jointcma}
\end{equation}

\textbf{Separable}: Similar as discussed in Section \ref{model:attention:sa}, we can reduce the space complexity of the intermediate representations of the proposed cross-modal attention layer by performing a separate self-attention across the time-axis for each modality individually and then perform CMA across the remaining axis (i.e.\ sources or spatial locations) as shown next, also illustrated in Figure \ref{fig:attention_architectures}(c):
\begin{equation}
    \begin{aligned}
        a^{(l)} = & \operatorname{SA}_{\{\mathrm{T}\}}(A_\mathrm{CMA}^{(l-1)}) \\
        v^{(l)} = & \operatorname{SA}_{\{\mathrm{T}\}}(V_\mathrm{CMA}^{(l-1)}) \\
        A_\mathrm{CMA}^{(l)}, V_\mathrm{CMA}^{(l)} = & \operatorname{CMA}_{\{\mathrm{M|G}\}}(a^{(l)}, v^{(l)}).
    \end{aligned}
    \label{eq:separable_cma}
\end{equation}

\subsection{Audio-visual on-screen sound classifier}
\label{model:classifier}
The audio-visual correspondence between each estimated source $\hat{s}_m$ and the video is computed using the extracted audio-visual representation from the output of our transformer based models defined in Equations (\ref{eq:final_jointselfattetnion}) and (\ref{eq:final_jointcma}), for the self-attention and cross-modal attention encoders, respectively. Specifically, we feed this audio-visual embedding $z \in \R^{M \times D}$ through a dense layer $f_z$ and we apply a sigmoid activation element-wise to compute the audio-visual coincidence probabilities $\hat{y}_{1:M}$ for all sources. Therefore, the final on-screen waveform estimate $\hat{x}^\mathrm{on}$ is produced using these probabilities as soft weights and multiplied with the corresponding estimated sources:
\begin{equation}
    \hat{y}_m = \sigma \left( f_z (z) \right)_m \in [0, 1],  \forall \enskip m = {1,\dots, M}, \enskip 
    \label{eq:onscreen_estimate}
    \hat{x}^\mathrm{on} =
    \sum_{m=1}^{M}
    \hat{y}_m \hat{s}_m.
\end{equation}

\section{Experimental Framework}

\subsection{Data preparation}

We use the same training data as AudioScope \cite{tzinis2021into}, except instead of only using a subset of the data filtered by an audio-visual coincidence prediction model \cite{jansen2020coincidence}, we use all data, respecting the published splits\footnote{\url{https://github.com/google-research/sound-separation/tree/master/datasets/yfcc100m}} for the Yahoo Flickr Creative Commons 100 Million Dataset (YFCC100m) \cite{thomee2016yfcc100m}. The unfiltered training data consists of about 1600 hours. We gathered human annotations for 5 second clips from unfiltered videos from the validation and test splits. Out of 6500 validation clips labeled, 109 were unanimously rated as on-screen-only, and 1421 unanimous off-screen-only. For 3500 test clips, there were 44 unanimous on-screen-only, and 788 unanimous off-screen-only. We also experiment with using a faster video frame rate of 16 frames per second (FPS), instead of 1 FPS as used by AudioScope \cite{tzinis2021into}.

\subsection{Training}
\label{ssec:training}

We use the same training procedure as AudioScope \cite{tzinis2021into}. We focus on the unsupervised scenario, which means all batches are composed of ``noisy-labeled on-screen (NOn)'' examples (in the terminology of the original paper \cite{tzinis2021into}). Each NOn example consists of the video frames and audio for a 5 second video clip, where additional audio from another random 5 second video clip is used as synthetic off-screen audio, and is added to the audio of the first clip. These examples provide noisy labels, because we make the assumption that all sources that map to the first clip's audio are on-screen. However, not all sources that map to the clip's audio will necessarily be on-screen (e.g. there could be off-screen background noise in the first clip that is separated as a source).

Despite the training examples having these noisy labels, we found that the exact cross-entropy loss works well, which is computed between the pseudo-label assignments $y$ provided by MixIT and the audio-visual on-screen classifier predictions $\hat{y}$ provided by the model. Both audio and visual embedding networks were pre-trained on AudioSet \cite{AudioSet} for unsupervised coincidence prediction \cite{jansen2020coincidence}. But unlike AudioScope, we freeze these networks during training, which we found to yield better results. Also, instead of initializing the separation model from scratch, we initialize the separation model with weights pre-trained on unfiltered audio-only MoMs drawn from YFCC100m, trained for 3.6M steps, which also significantly boosted the performance of our models. All models are trained on 32 Google Cloud TPU v3 cores with the Adam optimizer \cite{kingma2014adam}, a batch size of $256$, and learning rate of $10^{-4}$.

\subsubsection{Calibration}

In contrast to AudioScope, we perform an additional post-training calibration step. Using validation data that is human-labeled as unanimously on-screen-only or unanimously off-screen-only, we calibrate the on-screen classifier using the \texttt{scikit-learn} \cite{scikit-learn} implementation of isotonic regression \cite{zadrozny2002transforming}. The dataset for this calibration consists of both single-mixture and MoM examples. For single-mixture examples, the labels are all 1 for on-screen-only videos, and all 0 for off-screen only videos. MoM examples are constructed from an on-screen-only or off-screen-only video and the audio from either off-screen-only videos, or random videos chosen from the entire validation set. For on-screen MoM examples, the labels are the MixIT assignments for the on-screen mixture, so they are 1 for sources that map to the on-screen-only video, and 0 otherwise. For off-screen MoM examples, the labels are all 0.

\subsection{Evaluation}
\vspace{-5pt}

For AudioScope \cite{tzinis2021into}, the videos in the validation and test sets were drawn from a subset of YFCC100m filtered by an unsupervised coincidence model \cite{jansen2020coincidence}. MoM examples in this dataset use unanimously-rated off-screen-only videos as the background regardless of whether the foreground video is on-screen-only or off-screen-only. We refer to this as ``filtered off-screen background.'' In addition to this filtered dataset, we constructed new test sets drawn from the unfiltered validation and test splits of YFCC100m. We created two versions of these unfiltered datasets: one version, ``unfiltered off-screen background,'' uses audio from unanimously-rated off-screen-only unfiltered videos as the background in MoMs, and the other version, ``unfiltered random background,'' uses audio sampled from all videos in the unfiltered validation and test sets.

Evaluation metrics are the same as used for AudioScope \cite{tzinis2021into}: we measure power-weighted area under the curve of the receiver operating characteristic (AUC-ROC), scale-invariant signal-to-noise ratio (SI-SNR) \cite{LeRoux2018a} of the on-screen estimate $\hat{x}^\mathrm{on}$ for on-screen examples (a measure of the reconstruction fidelity of on-screen audio), and off-screen suppression ratio (OSR) for off-screen examples, which measures the power reduction of the on-screen estimate $\hat{x}^\mathrm{on}$ relative to the input audio mixture power. These metrics are measured on both single mixtures and MoMs. For MoMs, we additionally report an oracle metric, MixIT$^*$, which is the SI-SNR of the estimated on-screen audio using the MixIT assignments from separated sources to the reference on-screen audio.

\vspace{-5pt}
\section{Results}
\label{results}
\vspace{-5pt}

Results are shown in Tables \ref{tab:results_f_off}, \ref{tab:results_unf_off}, and \ref{tab:results_unf_random} for filtered off-screen background, unfiltered off-screen background, and unfitered random background, respectively. The first row in each table lists the performance of the equivalent original AudioScope \cite{tzinis2021into} model. Note that this model performs relatively well on the filtered off-screen background dataset (Table \ref{tab:results_f_off}), but its performance drops precipitously on the unfiltered datasets (Tables \ref{tab:results_unf_off} and \ref{tab:results_unf_random}). 
In particular, AUC-ROC drops from from 0.81 to 0.63 and 0.65, and on-screen MoM SI-SNR of $\hat{x}^\mathrm{on}$ drops from 8.0 dB to 0.7 dB and -0.7 dB. Notice that the OSRs actually increase from filtered to unfiltered evaluation; this is due to the model tending to predict probability 0 for all sources regardless of whether the video is on-screen or not. Thus, AudioScope exhibits significant mismatch to the unfiltered evaluation datasets.

Using a pre-trained separation model and fine-tuning on unfiltered boosts the performance of AudioScope by about 2 dB in terms of both oracle MixIT* on-screen SI-SNR and $\hat{x}^\mathrm{on}$ on-screen SI-SNR, and also generalizes better to unfiltered datasets. There is a slight performance improvement using 16 FPS; in particular, OSR increases a bit.
Notice that using calibration shifts the operating point of models: it tends to reduce on-screen SI-SNR by up to 1 dB, while substantially boosting OSR. This is likely because the uncalibrated on-screen probablities tend towards predicting 1, and this bias degrades rejection of off-screen sounds when the on-screen estimate is created.

Our proposed attention-based models further improve performance over pre-trained models that use the original spatio-temporal attention used by AudioScope \cite{tzinis2021into}. 
For calibrated models, on filtered off-screen background MoMs, our attention-based models achieve a slight gain of 0.4 dB on-screen SI-SNR and 0.6 dB OSR. On unfiltered MoMs with off-screen background, attention-based models boost on-screen SI-SNR by 1.2 dB with comparable OSR, and on unfiltered random background MoMs gain almost 3 dB on-screen SI-SNR, with only slightly lower OSR. 
It is remarkable that the performance improvements for these attention based models seem to increase as the difficulty of the evaluation data increases.

\begin{table}[!h]
\caption{Evaluation results for ``filtered off-screen background'' test set with calibration. On-screen MoMs have a median input SI-SNR of 4.4 dB. ``PT'' indicates separation model pre-training, and ``Cal.'' indicates calibration on labeled validation data.}
\label{tab:results_f_off}
\vspace{-5pt}
\begin{center}
\scalebox{0.72}{
\begin{tabular}{lllrrccrrrc}
\toprule
&&&
&\multicolumn{3}{c}{\bf Single mixture}
&\multicolumn{4}{c}{\bf Mixture of mixtures}
\\
\cmidrule(lr){5-7} \cmidrule(lr){8-11}
&&&&
&\multicolumn{1}{c}{\bf SI-SNR (dB)}
&\multicolumn{1}{c}{\bf OSR (dB)}
&
&\multicolumn{2}{c}{\bf SI-SNR (dB)}
&\multicolumn{1}{c}{\bf OSR (dB)}
\\
\cmidrule(lr){6-6} \cmidrule(lr){7-7}
\cmidrule(lr){9-10} \cmidrule(lr){11-11}
\multicolumn{1}{c}{\bf AV alignment} 
&\multicolumn{1}{c}{\bf PT} 
&\multicolumn{1}{c}{\bf Cal.} 
&\multicolumn{1}{c}{\bf Train data}
&\multicolumn{1}{c}{\bf AUC}
&\multicolumn{1}{c}{\bf On: $\hat{x}^\mathrm{on}$}
&\multicolumn{1}{c}{\bf Off: $\hat{x}^\mathrm{on}$}
&\multicolumn{1}{c}{\bf AUC}
&\multicolumn{1}{c}{\bf MixIT*}
&\multicolumn{1}{c}{\bf On: $\hat{x}^\mathrm{on}$}
&\multicolumn{1}{c}{\bf Off: $\hat{x}^\mathrm{on}$}
\\
\midrule
                          Spatio-temporal \cite{tzinis2021into} & & & Filt.\ 1 FPS &                        0.62 &                                     \textbf{36.6} &                                                0.5 &                                         0.81 &                           10.6 &                                                                8.0 &                                                                 5.3 \\
\midrule
                                  Spatio-temporal & \Checkmark & & Unfil.\ 1 FPS &                        0.58 &                                              29.2 &                                                1.7 &                                         0.79 &                           12.5 &                                                                9.7 &                                                                 4.0 \\
                                 Spatio-temporal & \Checkmark & & Unfil.\ 16 FPS &                        0.65 &                                              29.0 &                                                1.6 &                                         0.82 &                  12.5 &                                                               10.0 &                                                                 3.7 \\
                       Spatio-temporal & \Checkmark & \Checkmark & Unfil.\ 1 FPS &                        0.58 &                                              33.0 &                                       \textbf{9.5} &                                         0.79 &                           12.5 &                                                                9.4 &                                                                10.2 \\
                      Spatio-temporal & \Checkmark & \Checkmark & Unfil.\ 16 FPS &                        0.65 &                                              30.6 &                                                9.2 &                                         0.82 &                  12.5 &                                                                8.4 &                                                                10.2 \\
\midrule
                              Joint SA $\times4$ & \Checkmark & & Unfil.\ 16 FPS &                        0.67 &                                              29.4 &                                                1.5 &                                         0.84 &                           12.5 &                                                               \textbf{11.0} &                                                                 4.6 \\
                          Separable SA $\times4$ & \Checkmark & & Unfil.\ 16 FPS &                        \textbf{0.71} &                                              28.8 &                                                2.2 &                                         0.85 &                           12.5 &                                                               10.9 &                                                                 4.9 \\
                             Joint CMA $\times4$ & \Checkmark & & Unfil.\ 16 FPS &               \textbf{0.71} &                                              30.8 &                                                1.8 &                                \textbf{0.86} &                           12.5 &                                                      \textbf{11.0} &                                                                 4.7 \\
                         Separable CMA $\times4$ & \Checkmark & & Unfil.\ 16 FPS &                        0.69 &                                              33.2 &                                                1.3 &                                         0.85 &                           12.5 &                                                               10.9 &                                                                 4.3 \\
                   Joint SA $\times4$ & \Checkmark & \Checkmark & Unfil.\ 16 FPS &                        0.67 &                                              31.3 &                                                8.6 &                                         0.84 &                           12.5 &                                                                9.5 &                                                                10.6 \\
               Separable SA $\times4$ & \Checkmark & \Checkmark & Unfil.\ 16 FPS &                        \textbf{0.71} &                                              30.4 &                                                8.9 &                                         0.85 &                           12.5 &                                                                9.1 &                                                                10.7 \\
                  Joint CMA $\times4$ & \Checkmark & \Checkmark & Unfil.\ 16 FPS &               \textbf{0.71} &                                              30.5 &                                                8.9 &                                \textbf{0.86} &                           12.5 &                                                                9.8 &                                                       \textbf{10.8} \\
              Separable CMA $\times4$ & \Checkmark & \Checkmark & Unfil.\ 16 FPS &                        0.69 &                                              32.5 &                                                8.1 &                                         0.85 &                           12.5 &                                                                9.6 &                                                                10.6 \\

\bottomrule
\end{tabular}
}
\end{center}
\vspace{-10pt}
\end{table}

\begin{table}[!h]
\caption{Evaluation results for ``unfiltered off-screen background'' test set with calibration. On-screen MoMs have a median input SI-SNR of 4.0 dB.}
\label{tab:results_unf_off}
\vspace{-5pt}
\begin{center}
\scalebox{0.72}{
\begin{tabular}{llrrrccrrrc}
\toprule
&&&
&\multicolumn{3}{c}{\bf Single mixture}
&\multicolumn{4}{c}{\bf Mixture of mixtures}
\\
\cmidrule(lr){5-7} \cmidrule(lr){8-11}
&&&&
&\multicolumn{1}{c}{\bf SI-SNR (dB)}
&\multicolumn{1}{c}{\bf OSR (dB)}
&
&\multicolumn{2}{c}{\bf SI-SNR (dB)}
&\multicolumn{1}{c}{\bf OSR (dB)}
\\
\cmidrule(lr){6-6} \cmidrule(lr){7-7}
\cmidrule(lr){9-10} \cmidrule(lr){11-11}
\multicolumn{1}{c}{\bf AV alignment} 
&\multicolumn{1}{c}{\bf PT} 
&\multicolumn{1}{c}{\bf Cal.} 
&\multicolumn{1}{c}{\bf Train data}
&\multicolumn{1}{c}{\bf AUC}
&\multicolumn{1}{c}{\bf On: $\hat{x}^\mathrm{on}$}
&\multicolumn{1}{c}{\bf Off: $\hat{x}^\mathrm{on}$}
&\multicolumn{1}{c}{\bf AUC}
&\multicolumn{1}{c}{\bf MixIT*}
&\multicolumn{1}{c}{\bf On: $\hat{x}^\mathrm{on}$}
&\multicolumn{1}{c}{\bf Off: $\hat{x}^\mathrm{on}$}
\\
\midrule
                          Spatio-temporal \cite{tzinis2021into} & & & Filt.\ 1 FPS &                        0.54 &                                              19.5 &                                                9.1 &                                         0.63 &                           10.2 &                                                                0.7 &                                                                16.6 \\
\midrule
                                  Spatio-temporal & \Checkmark & & Unfil.\ 1 FPS &                        0.59 &                                              29.1 &                                                5.2 &                                         0.72 &                  12.1 &                                                                8.1 &                                                                 7.0 \\
                                 Spatio-temporal & \Checkmark & & Unfil.\ 16 FPS &                        0.70 &                                              30.2 &                                                4.9 &                                         0.78 &                           12.1 &                                                                8.4 &                                                                 6.3 \\
                       Spatio-temporal & \Checkmark & \Checkmark & Unfil.\ 1 FPS &                        0.59 &                                              30.2 &                                               22.6 &                                         0.72 &                  12.1 &                                                                8.6 &                                                                24.9 \\
                      Spatio-temporal & \Checkmark & \Checkmark & Unfil.\ 16 FPS &                        0.70 &                                              29.2 &                                      \textbf{24.8} &                                         0.78 &                           12.1 &                                                                8.4 &                                                                26.0 \\
\midrule
                              Joint SA $\times4$ & \Checkmark & & Unfil.\ 16 FPS &                        0.68 &                                              30.5 &                                                5.0 &                                         0.80 &                           12.0 &                                                               10.2 &                                                                 6.3 \\
                          Separable SA $\times4$ & \Checkmark & & Unfil.\ 16 FPS &               \textbf{0.77} &                                              30.4 &                                                5.2 &                                \textbf{0.83} &                           12.0 &                                                      \textbf{10.4} &                                                                 6.8 \\
                             Joint CMA $\times4$ & \Checkmark & & Unfil.\ 16 FPS &                        0.70 &                                              31.1 &                                                4.7 &                                         0.80 &                           12.0 &                                                                9.8 &                                                                 6.6 \\
                         Separable CMA $\times4$ & \Checkmark & & Unfil.\ 16 FPS &                        0.69 &                                              32.1 &                                                4.1 &                                         0.81 &                           12.1 &                                                                9.8 &                                                                 6.4 \\
                   Joint SA $\times4$ & \Checkmark & \Checkmark & Unfil.\ 16 FPS &                        0.68 &                                              29.9 &                                               24.6 &                                         0.80 &                           12.0 &                                                                9.1 &                                                                25.1 \\
               Separable SA $\times4$ & \Checkmark & \Checkmark & Unfil.\ 16 FPS &               \textbf{0.77} &                                              29.8 &                                               \textbf{24.8} &                                \textbf{0.83} &                           12.0 &                                                                9.8 &                                                       \textbf{26.2} \\
                  Joint CMA $\times4$ & \Checkmark & \Checkmark & Unfil.\ 16 FPS &                        0.70 &                                     \textbf{32.1} &                                               24.5 &                                         0.80 &                           12.0 &                                                                9.5 &                                                                25.7 \\
              Separable CMA $\times4$ & \Checkmark & \Checkmark & Unfil.\ 16 FPS &                        0.69 &                                              32.0 &                                               23.1 &                                         0.81 &                           12.1 &                                                                9.0 &                                                                25.2 \\

\bottomrule
\end{tabular}
}
\end{center}
\vspace{-10pt}
\end{table}
\begin{table}[!h]
\caption{Evaluation results for ``unfiltered random background'' test set with calibration. On-screen MoMs have a median input SI-SNR of 2.5 dB.}
\label{tab:results_unf_random}
\vspace{-5pt}
\begin{center}
\scalebox{0.72}{
\begin{tabular}{llrrrccrrrc}
\toprule
&&&
&\multicolumn{3}{c}{\bf Single mixture}
&\multicolumn{4}{c}{\bf Mixture of mixtures}
\\
\cmidrule(lr){5-7} \cmidrule(lr){8-11}
&&&&
&\multicolumn{1}{c}{\bf SI-SNR (dB)}
&\multicolumn{1}{c}{\bf OSR (dB)}
&
&\multicolumn{2}{c}{\bf SI-SNR (dB)}
&\multicolumn{1}{c}{\bf OSR (dB)}
\\
\cmidrule(lr){6-6} \cmidrule(lr){7-7}
\cmidrule(lr){9-10} \cmidrule(lr){11-11}
\multicolumn{1}{c}{\bf AV alignment} 
&\multicolumn{1}{c}{\bf PT} 
&\multicolumn{1}{c}{\bf Cal.} 
&\multicolumn{1}{c}{\bf Train data}
&\multicolumn{1}{c}{\bf AUC}
&\multicolumn{1}{c}{\bf On: $\hat{x}^\mathrm{on}$}
&\multicolumn{1}{c}{\bf Off: $\hat{x}^\mathrm{on}$}
&\multicolumn{1}{c}{\bf AUC}
&\multicolumn{1}{c}{\bf MixIT*}
&\multicolumn{1}{c}{\bf On: $\hat{x}^\mathrm{on}$}
&\multicolumn{1}{c}{\bf Off: $\hat{x}^\mathrm{on}$}
\\
\midrule
                          Spatio-temporal \cite{tzinis2021into} & & & Filt.\ 1 FPS &                        0.54 &                                              19.5 &                                                9.1 &                                         0.65 &                            8.0 &                                                               -0.7 &                                                                17.4 \\
\midrule
                                  Spatio-temporal & \Checkmark & & Unfil.\ 1 FPS &                        0.59 &                                              29.1 &                                                5.2 &                                         0.72 &                           10.1 &                                                                5.9 &                                                                 5.6 \\
                                 Spatio-temporal & \Checkmark & & Unfil.\ 16 FPS &                        0.70 &                                              30.2 &                                                4.9 &                                         0.73 &                           10.1 &                                                                5.8 &                                                                 4.2 \\
                       Spatio-temporal & \Checkmark & \Checkmark & Unfil.\ 1 FPS &                        0.59 &                                              30.2 &                                               22.6 &                                         0.72 &                           10.1 &                                                                5.8 &                                                       \textbf{23.4} \\
                      Spatio-temporal & \Checkmark & \Checkmark & Unfil.\ 16 FPS &                        0.70 &                                              29.2 &                                      \textbf{24.8} &                                         0.73 &                           10.1 &                                                                5.2 &                                                                22.5 \\
\midrule
                              Joint SA $\times4$ & \Checkmark & & Unfil.\ 16 FPS &                        0.68 &                                              30.5 &                                                5.0 &                                         0.76 &                           10.1 &                                                       \textbf{8.9} &                                                                 3.9 \\
                          Separable SA $\times4$ & \Checkmark & & Unfil.\ 16 FPS &               \textbf{0.77} &                                              30.4 &                                                5.2 &                                \textbf{0.78} &                           10.1 &                                                                \textbf{8.9} &                                                                 4.1 \\
                             Joint CMA $\times4$ & \Checkmark & & Unfil.\ 16 FPS &                        0.70 &                                              31.1 &                                                4.7 &                                         0.75 &                  10.2 &                                                                8.0 &                                                                 4.4 \\
                         Separable CMA $\times4$ & \Checkmark & & Unfil.\ 16 FPS &                        0.69 &                                              32.1 &                                                4.1 &                                         0.77 &                           10.1 &                                                                \textbf{8.9} &                                                                 3.9 \\
                   Joint SA $\times4$ & \Checkmark & \Checkmark & Unfil.\ 16 FPS &                        0.68 &                                              29.9 &                                               24.6 &                                         0.76 &                           10.1 &                                                                8.1 &                                                                21.2 \\
               Separable SA $\times4$ & \Checkmark & \Checkmark & Unfil.\ 16 FPS &               \textbf{0.77} &                                              29.8 &                                               \textbf{24.8} &                                \textbf{0.78} &                           10.1 &                                                                8.4 &                                                                20.0 \\
                  Joint CMA $\times4$ & \Checkmark & \Checkmark & Unfil.\ 16 FPS &                        0.70 &                                     \textbf{32.1} &                                               24.5 &                                         0.75 &                  10.2 &                                                                7.1 &                                                                20.6 \\
              Separable CMA $\times4$ & \Checkmark & \Checkmark & Unfil.\ 16 FPS &                        0.69 &                                              32.0 &                                               23.1 &                                         0.77 &                           10.1 &                                                                8.5 &                                                                20.6 \\
\bottomrule
\end{tabular}
}
\end{center}
\end{table}

\vspace{-5pt}
\section{Conclusion}
\label{conclusion}
\vspace{-5pt}

In this paper we have presented extensions and refinements of the AudioScope model
that further improve upon previous audio-visual self-supervised methods for separating on-screen sounds. Our proposed model is able to operate on higher frame-rates and build rich semantic associations between audio and video modalities using the proposed self-attention as well as cross-modal attention mechanisms. Our experimental results show that our model is able to generalize to a much wider set of in-the-wild videos than existing approaches while being trained solely with in-the-wild videos.  The approach we followed here is to first separate the audio before aligning each on-screen source with its corresponding video object.  Other works have followed the converse process of first identifying visual objects and using them to condition sound separation for each object.  Future works may explore unifying these two directions into a single model that works in both ways.

\bibliographystyle{abbrv}
\bibliography{refs}

\end{document}